# STRUCTURAL ASPECTS OF THE METAL-INSULATOR TRANSITION IN BaVS$_3$

**S. Fagot, P. Foury-Leylekian, S. Ravy[a], J.P. Pouget**

Laboratoire de Physique des Solides, CNRS UMR 8502, Bât. 510, Université Paris Sud, 91405 Orsay cedex, France

[a] Synchrotron SOLEIL, Saint Aubin BP 48, 91192 Gif Sur Yvette, France

**M. Anne**

Laboratoire de Cristallographie, CNRS, Bâtiment F, 25 Av. des martyrs, 38042 Grenoble cedex 9, France

**G. Popov, M.V. Lobanov, M. Greenblatt**

Department of Chemistry and Chemical Biology, Rutgers, the State University of New Jersey, Piscataway, NJ 08854 USA

**Abstract**

A sequence of structural transitions occurring in the quasi-one-dimensional (1D) 3d$^1$ system BaVS$_3$ at low temperature was investigated by high resolution synchrotron X-ray diffraction. The orthorhombic *Cmc2$_1$* structure of the intermediate-temperature (70K<T<240K) phase was confirmed. A model for the low-T (T<70K) $k=(1\ 0\ ½)_O$ superstructure (with *Im* symmetry) is proposed and refined. The formation of the superstructure is associated with the stabilization of a mixed bond order / charge density wave.

**Introduction**

1. General features

Low-dimensional strongly correlated electronic systems are intensively studied because they exhibit a rich variety of quantum ground states such as superconductivity, charge and spin density waves, and orbital ordering[1]. Of particular interest is the metal-insulator (MI) Mott-Hubbard transition driven by electron-electron repulsions and the anomalous metallic properties associated with the low dimension of the electron gas. Transition metal oxides have been intensively studied since the discovery of high temperature superconductivity in competition with antiferromagnetism in cuprates, the observation of competing spin, charge, lattice and orbital interactions in manganites[2] and stabilization of numerous charge density waves (CDW) ground states in Mo and W bronzes[3].

In this context, the study of metallic sulphides is especially interesting. BaVS$_3$ is a quasi-1D metal which exhibits anomalous electronic and magnetic properties[4-8]. Furthermore, BaVS$_3$ has a unique electronic structure. Indeed, due to the distorted (D$_{3d}$) octahedral sulphur environment of the vanadium, low energy electronic states are formed by crystal-field split V(3d) orbitals: a higher energy $d_{z^2}$ level and two lower energy quasi-degenerate "$d_{xy}$" and "$d_{x^2-y^2}$"-like levels (called e($t_{2g}$) below)[8]. First principles electronic calculations[9,10] using linear augmented plane waves indicate that these three orbitals give rise to three bands crossing the Fermi level: a quasi 1D broad $d_{z^2}$ band, extended essentially along the c* direction of the hexagonal lattice and nearly filled by the two $3d^1$ electrons provided by the two $V^{4+}$ of the unit cell (the calculated Fermi wavevector is $k_{Fcalc}$=0.47c*) and two nearly non-dispersive e($t_{2g}$) bands that are almost empty. The $d_{z^2}$ band accounts for the metallic character of $BaVS_3$ while the e($t_{2g}$) localized levels are responsible for its magnetism[8].

$BaVS_3$ undergoes a remarkable succession of three phase transitions whose origin is still a matter of controversy. $BaVS_3$ crystallizes in the hexagonal *P6₃/mmc* space group with unit cell parameters: $a_H$~6.71 Å and $c_H$~5.62 Å at room temperature. Its structure consists of a slightly distorted hexagonal close-packed network of face sharing $VS_6$ octahedral chains directed along the *c* axis[4]. In the *(ab)* planes, the $V^{4+}$ chains are well separated by Ba atoms, leading to an interchain V-V distance of 6.71 Å, while the intrachain V-V distance is only 2.81 Å. At $T_S$= 240 K $BaVS_3$ undergoes a second order structural phase transition without any significant change in the electronic properties. The lattice symmetry becomes orthorhombic[11] with unit cell parameters: $a_O$~ 6.75 Å, $b_O$~ 11.48 Å $<\sqrt{3}\ a_O$ and $c_O$~ 5.60 Å at 100 K. This orthorhombic structure has been recently questioned[12], because refinements of powder neutron diffraction data performed in Ref. 11 could barely discriminate between three different space groups corresponding to different distortions of the $VS_6$ octahedral chains. In this paper, we show that the correct space group is *Cmc2₁*, in agreement with the conclusions of Ref. 11.

2. The metal-insulator transition of $BaVS_3$

At $T_{MI}$= 70 K, $BaVS_3$ undergoes a MI transition characterized by a strong increase of resistivity[6-8], with an activated thermal dependence leading to a charge gap of 2Δ~ 350K (Ref. 6) - 600K (Ref. 7) . $T_{MI}$ is also characterized by a change of slope in the thermal dependence of the lattice parameters[13] and a specific heat anomaly[14]. The magnetic properties are also significantly affected by the MI transition. Above $T_{MI}$ $BaVS_3$ is paramagnetic with dominant ferromagnetic interactions. Below $T_{MI}$ the magnetic susceptibility decreases rapidly but remains finite at low temperature. The spin gap is estimated to be ~120K-250K at low temperature[15]. The MI transition is suppressed by applying pressure[16,17].

Recent structural studies[12,18] have revealed that the MI transition is also accompanied by a structural transition evidenced by the appearance of superlattice reflections at the (1, 0, ½)$_O$ reduced wave vector with respect to the orthorhombic setting (T point of the Brillouin zone). At 20K the intensity of the satellite reflections is ~$10^{-2}$ of that of the main Bragg reflections.

The $T_{MI}$ transition bears some similarity with the Peierls transition observed in quasi-1D metals in which a $2k_F$ CDW is stabilized ($k_F$ is the Fermi wave vector of the 1D electron gas). This interpretation is sustained by the observation of 1D pretransitional fluctuations at the critical wave vector 0.5c* (ascribed to $2k_{Fexp}$ in Ref. 18) in an extended T range up to ~170K. The CDW scenario, developed in Ref. 18, relies on the presence of a half-filled 1D $d_{z^2}$ band, at variance with the results of standard first principle electronic calculations [9,10]. In this scenario, the MI transition is due to the opening of an energy gap mostly in the $dz^2$ band.

In addition, a significant change of about 7% in the intensity of the main Bragg reflection is also observed[18] at $T_{MI}$. This indicates quite a strong modification of the average crystallographic structure which is not preceded by pretransitional fluctuations. In this respect the MI transition does not appear to be a conventional Peierls transition. Several models have been proposed to explain these unusual structural features[18]. However, in order to discriminate between these scenarios, an accurate determination of the structure of the insulating phase of $BaVS_3$ is required.

### 3. The mysterious $T_x$ transition

At $T_x \sim 30K$, a weak change of slope in the thermal dependence of the resistivity and a strong anomaly in the $\chi_c - \chi_a$ anisotropy of the magnetic susceptibility[7] are detected. While NMR/NQR measurements have failed to detect any magnetic order below $T_x$[15], a recent neutron scattering study[19] has provided evidence for an incommensurate magnetic order with the propagation vector q= (0.226, 0.226, 0)$_H$ in the hexagonal setting. According to recent X-ray studies[12,18], no additional structural features have been detected at the $T_x$ phase transition.

It has been suggested that a pairing of antiferromagnetically coupled electrons occurs below $T_x$. Several models of pairing involving either intrachain or interchain $V^{4+}$- $V^{4+}$ dimers have been considered[20]. In these scenarios there is a coupling between the magnetic order and an orbital ordering. However, up to now there has not been any clear experimental evidence of an orbital ordering below $T_x$, except perhaps in a recent NQR study[15].

In this paper, we present a determination of the crystallographic structure of the two low temperature insulating phases of $BaVS_3$ from high resolution synchrotron powder X-ray diffraction.

**Experimental details**

The synthesis technique was similar to the one described in Ref. 21. At least 99.99% pure $BaCO_3$ and $V_2O_5$ were mixed, thoroughly ground, pressed into pellets and sintered in a flow of nitrogen bubbled through $CS_2$ liquid. The sample was sintered 5 hours at 500°C, 5 hours at 600°C and a total of 114 hours at 900°C with two intermediate grindings. The sample appears to be sulphur-stoichiometric according to magnetic susceptibility measurements, as is to be expected for the synthetic procedure used. High resolution synchrotron X-ray powder diffraction data were collected at the beam line ID31 at the ESRF. The powder diffraction patterns were collected at 300K, 100K, 40K and 5K under the same conditions. Continuous scans from 4° to 56° (2θ) were recorded in the Debye-Scherrer geometry during either 12h or 24h with a wavelength of 0.501813 Å using a 9-channel crystal analyser. The profile refinements were carried out with the Rietveld program FULLPROF[22]. The BaVS$_3$ powder had a grain size of ~1µm as determined by optical microscopy.

**Results and discussion**

1 The 300K and 100K crystallographic structures

The structural refinement performed at 300K confirms the *P6$_3$/mmc* hexagonal structure[4,11]. The profile parameters and the conventional R-factors are given in Table 1. The atomic coordinates are listed in Table 2a. The refinement revealed a large and anisotropic Debye-Waller factor for the vanadium atom ($B_{iso}$~0.04 Å$^2$), with the dominant component in the *ab* hexagonal plane ($U_{11}/U_{33}$~2.9). This shows that already at 300K the V atom, located at the center of the VS$_6$ octahedron, is subject to a structural instability, precursor of the $T_S$ phase transition.

Figure 1 shows the result of the Rietveld refinement of the orthorhombic structure at 100K. The systematic extinctions at *h+k* odd and (*h 0 l*), *l* odd lead to 3 possible space groups: *Cmcm*, *Cmc2$_1$* and *C2cm* (*Ama2* in the standard setting). Structural refinements were performed in these space groups. In Ref. 11 the space group *C222$_1$* was also considered; here, it has been discarded on the basis of the observed extinctions. The best refinement was obtained with the *Cmc2$_1$* model yielding $\chi^2$= 12.99, while the *Cmcm* model led to $\chi^2$= 15.72 and the *C2cm* one did not converge. In addition, the ratio of the R$_B$ factors for the *Cmcm* and *Cmc2$_1$* models R$_B$(*Cmcm*)/R$_B$(*Cmc2$_1$*)= 1.54 is significantly greater than the generalized R statistical factor estimated for this case (Hamilton test[23]). Moreover, the intensities of several reflections (for example the (331) and (061) reflections) can not be adequately fitted in the *Cmcm* model, while the *Cmc2$_1$* model allows their accurate description. The high $\chi^2$ value found is probably due to the asymmetry of the peak profile which cannot be correctly modeled by FULLPROF (see inset of Fig. 1) and to the high statistics of the data (low R$_{exp}$). This peak asymmetry has been ascribed to the presence of a non-uniform orthorhombic deformation. A possible explanation could be the formation of anisotropic microstrains at the ferroelastic *P6$_3$/mmc* → *Cmc2$_1$* structural phase transition.

The refinement parameters and atomic coordinates are given in Tables 1 and 2, respectively. Interatomic distances are indicated in Table 3 . Our results are in excellent agreement with those already published[11]. The uniform VS$_6$ octahedral chains along the *c* hexagonal direction transform into zigzag chains (running along the *2$_1$* screw axis) at the orthorhombic transition. At 100K the V atoms shift from the inversion center of the VS$_6$ octahedron in an alternate manner, zigzagging by ±0.256 Å along the b axis in the mirror plane (*m*) of the structure. This leads to a non-centrosymmetric orthorhombic structure. The orthorhombic transition results in a significant increase of the V-V intrachain distance from 2.808 Å (at 300K in the hexagonal phase) to 2.845 Å (at 100K). There are two chains per unit cell related by the *C* face centering; the zigzag displacements are in phase between these chains.

2 Low temperature superstructure

The powder diffraction pattern taken at 40K exhibits two remarkable features. A comparison with the 100K pattern reveals that several peaks are split (right inset of Fig. 2), which is the signature of a weak monoclinic distortion whose unique axis is the orthorhombic $a_O$ axis. This result corroborates the finding of a lowering of the Laue symmetry at $T_{MI}$ (Ref. 12). However, the magnitude of the monoclinic distortion is extremely small: $β_M$ (the angle between $b_O$ and $c_O$) deviates from 90° by only 0.05°. Additional superlattice reflections, which could be indexed on the basis of the [1, 0, ½]$_O$ propagation vector (left inset of Fig. 2) are also observed. In spite of the very high statistics of the data, these superlattice peaks are very weak (maximum of relative intensity : ~0.4%). However, the refinement of the superstructure has been attempted and the results are presented below.

The possible model of the superstructure can be constructed using simple group-theoretical considerations. The program "Isotropy" was used for the analysis and starting model generation[24]. The isotropy subgroups associated with the irreducible representations (irreps) at the T-point are *Cm* (*Im*, to be consistent with our monoclinic setting) associated with the $T_1T_3$ irrep (Miller-Love[25] notations are used here and subsequently), and *Cc* (*Ia*), associated with the $T_2T_4$ irrep. The latter space group gives, in addition to the general extinction condition for the I centred lattice h+k+l even, supplementary extinction conditions, (h 0 l) h,l even, which are not observed; consequently, the refinement has been performed in the model with a *2c* supercell and *Im* symmetry. The *Im* symmetry implies that the distortions of the two chains of the unit cell are out of phase. The refinement was limited to common isotropic thermal parameters due to convergence constraints.

The refinement parameters and atomic coordinates are summarized in Tables 1 and 2. Interatomic distances are given in Table 3. Figure 3a represents schematically the atomic displacements of the $VS_3$ chain of the 40K superstructure compared to the 100K positions. The key result is that the V atoms are significantly displaced from their "*Cmc2$_1$* positions" along the chain direction. There are four non-equivalent V sites and four V-V distances: V1-V2= 2.90 Å, V2-V3= 2.94 Å, V3-V4= 2.75 Å and V4-V1= 2.81 Å.

The sulphur network is also substantially modified compared to the 100K structure. The $S_3$ triangular faces shared between the octahedra become tilted with respect to the (*ab*) plane: in particular, the opposite faces of the V1$S_6$ and V3$S_6$ octahedra are tilted in opposite directions, while the opposite faces of the V2S6 and V4S6 octahedra remain nearly parallel. The bond valence sum (BVS) of each V atom has been calculated using the Zachariasen's method[26,27] with $R_0$ = 2.226 Å and the values are: 4.05±0.25 for V1, 4.01±0.25 for V2, 4.03±0.25 for V3 and 4.24±0.25 for V4.

Thus, within the uncertainties, there is no charge disproportionation associated with the 40K superstructure. The absolute values of BVS are slightly overestimated, because of the use of standard (RT) BVS parameters ($R_0$) for the low temperature phase without correction for the thermal contraction.

### 4 Thermal evolution of the superstructure: 5 K vs 40K

The X-ray diffraction pattern at 5K is nearly identical to that at 40K. Neither additional peak splittings nor extra superlattice reflections have been detected. Consequently, the superstructure refinement has been carried out with the same *Im* model. The refinement parameters and atomic coordinates are presented in Tables 1 and 2. The interatomic distances are given in Tables 3. The V-V distances, given in Fig. 3b, are nearly the same as those of the 40K structure. However, the S stacking is modified: the $S_3$ triangular faces shared between the octahedra along the chain direction are strongly distorted - either stretched or compressed.

The BVS calculations yielded the values of 3.83±0.25 for V1, 4.14±0.25 for V2, 4.31±0.25 for V3 and 4.13±0.25 for V4 (Fig. 3b), corresponding to the charge disproportionation of ~0.5 electrons between the V3 and V1 atoms, which sets a $2k_F$ CDW. The charge ordering manifests itself in the distortion of the $VS_6$ octahedra. The compression of the faces of the $V3S_6$ octahedron is associated with an excess of electrons in the environment of the V3 and thus with an increase of the effective charge of the V3, while the opposite situation is encountered with the stretching of the faces of the $V1S_6$ octahedron. It is worth noting that the BVS method probably overestimate the charge disproportionation but give a reliable indication of the charge distribution over the 4 inequivalent vanadium of the unit cell.

**Conclusions**

The sequence of structural and electronic transitions in $BaVS_3$ primarily affects the $V^{4+}$ metallic chains. This is the case for the $T_S$ orthorhombic transition, where we have confirmed that a substantial transverse shift of the V atoms from the inversion centers of the $VS_6$ octahedra occurs below $T_S$. This $Cmc2_1$ structural distortion leads to zigzag V chains, while the metallic character of the compound is preserved. The MI transition is driven by a V chain instability[18], which results in a doubling of the unit cell periodicity in the chain direction. In the Peierls-CDW scenario considered in Ref. 18, this instability occurs at the critical wave vector $2k_F$. In this study we have shown that below $T_{MI}$ the structure undergoes a monoclinic distortion, and that there are four inequivalent V sites in the low-T Im superstructure.

The modulation of the V distances in the superstructure can be described as a superposition of two displacement waves (bond order waves, BOW) that shift the vanadium atoms from their positions in the $Cmc2_1$ structure: a $4k_F=c^*$ sinusoidal modulation of amplitude ~0.025 Å and a $2k_F$ sinusoidal modulation of amplitude

~0.075 Å (~ three times larger than the $4k_F$ modulation), where only the V1 and V3 atoms are displaced (in opposite directions).

The $2k_F$ modulation is announced by an extended critical regime of 1D fluctuations; accordingly the order parameter (OP), associated with a $T_1T_3$ irrep, is the primary OP of the transition. On the other hand, the existence of the $4k_F$ ($k=0$) modulation component is primarily responsible for the profound change in intensity of the basic reflections. The $\Gamma$ ($k=0$) irrep compatible with the *Cm* symmetry is $\Gamma_3$. Accordingly, the structural transition could be regarded as driven by the coupled $T_1T_3 + \Gamma_3$ irreps. The corresponding Landau free energy expansion contains two third-order terms[25], which couple the two-dimensional primary OP belonging to $T_1T_3$ ($\eta_1, \eta_2$) and one-dimensional secondary OP belonging to $\Gamma_3$ ($\varepsilon$):

$$F_3 = a(\eta_1^2 - \eta_2^2)\varepsilon + b\eta_1\eta_2\varepsilon$$

Thus, the growth of $\eta$ below $T_{MI}$ should induce that of $\varepsilon$ and the thermal dependence of $\eta$ should be directly proportional to the square of $\varepsilon$. This interpretation is further confirmed by the experimental observation[18] below $T_{MI}$ where the excess of intensity of the Bragg reflections (proportional to $\varepsilon$) varies with the intensity of the superstructure satellite reflection (proportional to the square of $\eta$).

According to our powder diffraction data, the superstructure can be adequately described using the same *Im* model with the $c_o$ parameter doubled both at 40K (above the "mysterious" $T_x$ transition) and at 5K (below $T_x$). However, the V environments undergo substantial modification as T is lowered, which suggests a charge ordering in the V chains at 5K, as is effectively revealed by BVS calculations. To determine whether the charge ordering is accompanied by orbital ordering, further experiments, specifically, X-ray anomalous diffraction are underway.

**Acknowledgment**

This paper is written in memory of Professor E.F. Bertaut who has always manifested a constant interest in the structural investigations performed at Orsay in the field of the quasi-one dimensional conductors. We thank Dr. Michela Brunelli, the ESRF local contact, for the excellent management of the experiment and his help with the data analysis. The Orsay team thanks J. Rodriguez-Carvajal for very usefull discussions. The work at Rutgers University was supported by NSF-DMR-0233697.

Table Captions

Table 1: Space group, lattice parameters and X-ray powder reliability factors[22] at 300K, 100K, 40K and 5K (note that in the monoclinic setup $b_M \approx a_O$, $a_M \approx b_O$ and $c_M \approx 2c_O$). N is the number of independent observations and P is the number of fitted parameters.

Table 2: a) Atomic coordinates and isotropic displacement parameters (Å$^2$) at 300K, 100K, 40K and 5K. b) Anisotropic displacement parameters (Å$^2$) at 300K and 100K. No anisotropic displacement parameters have been used for the refinement at 40K and 5K due to numerical convergence limitations

Table 3: Selected interatomic distances (Å) in the BaVS$_3$ structure at different temperatures. The average of distances of the same type and standard deviation ($s = \sqrt{\dfrac{\sum_i (x_i - <x>)^2}{(n-1)}}$) are also given.

Figure Captions

Figure 1: Observed (o) and calculated X-ray powder diffraction patterns at 100K. The vertical ticks mark the Bragg reflection positions and the diagram at the bottom indicates the difference between the observed and calculated spectra. The inset shows the asymmetry of the peak profile.

Figure 2: Observed (o) and calculated X-ray powder diffraction patterns at 40K. The vertical ticks mark the Bragg reflection positions and the diagram at the bottom indicates the difference between the observed and calculated spectra. The two insets show some superlattice reflections and the peak splitting due to the monoclinic distortion, respectively.

Figure 3: a) Schematic representation of the VS$_3$ chain in the 40K superstructure. Filled circles represent V atoms, large empty circles - Ba, and small empty circles - S. The arrows indicate the atomic displacements (multiplied by 20) from their positions at 100K. The non-equivalent V atoms are labeled by different numbers.

b) Schematic representation of the same VS$_3$ chain in the 5K superstructure. The arrows indicate the atomic displacements (multiplied by 20) from their positions at 40K. The numbers associated with double arrows represent V-V distances. The numbers on the right of numbered V atoms are calculated BVS values.

Table 1

| Temperature | 300 K | 100 K | 40 K | 5 K |
|---|---|---|---|---|
| Crystal system | hexagonal | orthorhombic | monoclinic | monoclinic |
| Space group | $P6_3/mmc$ | $Cmc2_1$ | $Im$ | $Im$ |
| a (Å) | 6.713 (1) | 6.751 (1) | 11.458 (1) | 11.456 (1) |
| b (Å) |  | 11.485 (1) | 6.764 (1) | 6.764 (1) |
| c (Å) | 5.615 (1) | 5.597 (1) | 11.190 (1) | 11.188 (1) |
| β (°) |  |  | 90.045 (1) | 90.048 (1) |
| V (Å$^3$) | 219.209 (1) | 433.912 (1) | 867.216 (2) | 866.921 (2) |
| Z | 2 | 4 | 8 | 8 |
| 2θ range (°) | 4.0-56.0 | 4.0-56.0 | 4.0-56.0 | 4.0-56.0 |
| number of reflections | 521 | 1419 | 5345 | 5345 |
| N | 21729 | 23764 | 12464 | 12370 |
| P | 20 | 42 | 56 | 56 |
| C | 0 | 0 | 0 | 0 |
| N - P + C | 21709 | 23722 | 12408 | 12314 |
| $R_B$ | 5.85 | 3.52 | 2.74 | 3.64 |
| $R_{wp}$ | 15.5 | 13.2 | 10.6 | 10.7 |
| $R_{exp}$ | 12.2 | 3.67 | 2.73 | 3.30 |
| chi2 | 1.620 | 12.99 | 15.15 | 10.42 |

Table 2 a

| atom | site | x | y | z | U iso |
|---|---|---|---|---|---|
| 300 K | | | | | |
| Ba | 2d | 1/3 | 2/3 | ¾ | 0.0194 (2) |
| V | 2a | 0 | 0 | 0 | 0.0369 (7) |
| S | 6h | 0.1648 (2) | 0.3297(2) | ¼ | 0.0178 (3) |
| | | | | | |
| 100 K | | | | | |
| Ba | 4a | 0 | 0.3364 | 0.25 | 0.0076 (1) |
| V | 4a | 0 | 0.0223 (1) | -0.0102 (3) | 0.0062 (5) |
| S1 | 4a | 0 | 0.8306 (1) | 0.2377 (9) | 0.0093 (7) |
| S2 | 8b | 0.2442 (2) | 0.0838 (1) | 0.2568 (8) | 0.0073 (4) |
| | | | | | |
| 40 K | | | | | |
| Ba1 | 2a | 0.3404 (9) | 0 | 0.1213 (5) | 0.0042 (1) |
| Ba2 | 2a | -0.3339 (3) | 0 | 0.3737 (4) | 0.0042 (1) |
| Ba3 | 2a | 0.3364 (1) | 0 | 0.625 | 0.0042 (1) |
| Ba4 | 2a | -0.3403 (7) | 0 | 0.8717 (5) | 0.0042 (1) |
| V1 | 2a | -0.0248 (9) | 0 | 0.2364 (6) | 0.0042 (1) |
| V2 | 2a | 0.0241 (9) | 0 | 0.4903 (11) | 0.0042 (1) |
| V3 | 2a | -0.0210 (9) | 0 | 0.7490 (8) | 0.0042 (1) |
| V4 | 2a | 0.0240 (7) | 0 | 0.9907 (10) | 0.0042 (1) |
| S11 | 2a | 0.8309 (14) | 0 | 0.1104 (11) | 0.0042 (1) |
| S12 | 2a | -0.8297 (11) | 0 | 0.3633 (15) | 0.0042 (1) |
| S13 | 2a | 0.8242 (11) | 0 | 0.6241 (14) | 0.0042 (1) |
| S14 | 2a | -0.8290 (15) | 0 | 0.8711 (15) | 0.0042 (1) |
| S21 | 4b | 0.0874 (11) | 0.2395 (26) | 0.1264 (11) | 0.0042 (1) |
| S22 | 4b | -0.0806 (13) | -0.2453 (24) | 0.3769 (13) | 0.0042 (1) |
| S23 | 4b | 0.0839 (15) | 0.2408 (31) | 0.6279 (10) | 0.0042 (1) |
| S24 | 4b | -0.0794 (11) | -0.2497 (17) | 0.8761 (13) | 0.0042 (1) |
| | | | | | |
| 5 K | | | | | |
| Ba1 | 2a | 0.3364 | 0 | 0.125 | 0.0037 (1) |
| Ba2 | 2a | -0.3380 (6) | 0 | 0.3781 (4) | 0.0037 (1) |
| Ba3 | 2a | 0.3325 (7) | 0 | 0.6291 (5) | 0.0037 (1) |
| Ba4 | 2a | -0.3440 (3) | 0 | 0.8755 (5) | 0.0037 (1) |
| V1 | 2a | -0.0281 (10) | 0 | 0.2397 (8) | 0.0037 (1) |
| V2 | 2a | 0.0208 (8) | 0 | 0.4935 (9) | 0.0037 (1) |
| V3 | 2a | -0.0272 (8) | 0 | 0.7533 (6) | 0.0037 (1) |
| V4 | 2a | 0.0194 (8) | 0 | 0.9945 (10) | 0.0037 (1) |
| S11 | 2a | 0.8235 (13) | 0 | 0.1130 (9) | 0.0037 (1) |
| S12 | 2a | -0.8312 (15) | 0 | 0.3697 (14) | 0.0037 (1) |
| S13 | 2a | 0.8252 (17) | 0 | 0.6299 (11) | 0.0037 (1) |
| S14 | 2a | -0.8351 (12) | 0 | 0.8734 (15) | 0.0037 (1) |
| S21 | 4b | 0.0831 (12) | 0.2429 (24) | 0.1318 (11) | 0.0037 (1) |
| S22 | 4b | -0.0857 (12) | -0.2459 (22) | 0.3817 (11) | 0.0037 (1) |
| S23 | 4b | 0.0797 (13) | 0.2354 (16) | 0.6317 (11) | 0.0037 (1) |
| S24 | 4b | -0.0820 (9) | -0.2515 (16) | 0.8798 (12) | 0.0037 (1) |

Table 2 b

| atom | $U_{11}$ | $U_{22}$ | $U_{33}$ | $U_{12}$ | $U_{13}$ | $U_{23}$ |
|---|---|---|---|---|---|---|
| 300 K | | | | | | |
| Ba | 0.0174(1) | 0.0174(1) | 0.0233(2) | 0.0087(1) | 0 | 0 |
| V | 0.0471(7) | 0.0471(7) | 0.0164(6) | 0.0235(7) | 0 | 0 |
| S | 0.0157(3) | 0.0157(3) | 0.0220(4) | 0.0079(3) | 0 | 0 |
| | | | | | | |
| 100 K | | | | | | |
| Ba | 0.0072(1) | 0.0080(1) | 0.0077(1) | 0 | 0 | 0.0005(3) |
| V | 0.0067(4) | 0.0048(5) | 0.0072(5) | 0 | 0 | 0.0019(6) |
| S1 | 0.0106(5) | 0.0087(5) | 0.0085(10) | 0 | 0 | -0.0014(10) |
| S2 | 0.0064(3) | 0.0081(3) | 0.0074(6) | -0.0019(3) | -0.0014(10) | 0.0024(8) |

Table 3

|  | 300 K |  | 100 K |  |
|---|---|---|---|---|
|  | V-V | 2.8077 (10) | V-V | 2.8450 (24) |
|  | averages |  |  |  |
|  | V-S | 2.3763 (10) | V-S1 | 2.6029 (32) |
|  |  |  | V-S1 | 2.2011 (37) |
|  |  |  | V-S2 | 2.3345 (31) x 2 |
|  |  |  | V-S2 | 2.4301 (27) x 2 |
|  | averages |  |  | 2.3889 |
|  |  |  |  | 0.1344 |
|  | S-S | 3.3207 (15) x 6 | S2-S2 | 3.2972 (16) x 2 |
|  |  |  | S1-S2 | 3.3452 (17) x 4 |
|  | S-S | 3.3999 (10) x 6 | S1-S2 | 3.4818 (55) x 2 |
|  |  |  | S1-S2 | 3.3059 (54) x 2 |
|  |  |  | S2-S2 | 3.3971 (50) x 2 |
|  | averages | 3.3603 |  | 3.3621 |
|  |  | 0.0414 |  | 0.0654 |

| 40 K |  |  |  |  |  |  |  |  |
|---|---|---|---|---|---|---|---|---|
| V1-V2 | 2.896 (14) |  |  |  |  |  |  |  |
| V2-V3 | 2.941 (15) |  |  |  |  |  |  |  |
| V3-V4 | 2.753 (14) |  |  |  |  |  |  |  |
| V4-V1 | 2.806 (14) |  |  |  |  |  |  |  |
| average | 2.8490 |  |  |  |  |  |  |  |
| s | 0.0853 |  |  |  |  |  |  |  |
|  | vanadium 1 |  | vanadium 2 |  | vanadium 3 |  | vanadium 4 |  |
| V1-S12 | 2.647 (17) | V2-S13 | 2.738 (17) | V3-S14 | 2.588 (20) | V4-S11 | 2.587 (18) |  |
| V1-S11 | 2.172 (17) | V2-S12 | 2.197 (18) | V3-S13 | 2.258 (17) | V4-S14 | 2.151 (20) |  |
| V1-S22 | 2.374 (16) x 2 | V2-S23 | 2.344 (19) x 2 | V3-S24 | 2.307 (14) x 2 | V4-S21 | 2.335 (17) x 2 |  |
| V1-S21 | 2.407 (16) x 2 | V2-S22 | 2.408 (17) x 2 | V3-S23 | 2.436 (19) x 2 | V4-S24 | 2.429 (15) x 2 |  |
| average | 2.3969 |  | 2.40642 |  | 2.3888 |  | 2.378 |  |
| s | 0.1514 |  | 0.17962 |  | 0.1223 |  | 0.1441 |  |
| S21-S21 | 3.239 (25) | S23-S23 | 3.257 (30) | S23-S23 | 3.257 (30) | S21-S21 | 3.239 (25) |  |
| S22-S22 | 3.318 (23) | S22-S22 | 3.318 (23) | S24-S24 | 3.377 (16) | S24-S24 | 3.377 (16) |  |
| S11-S21 | 3.360 (19) x 2 | S13-S23 | 3.393 (21) x 2 | S13-S23 | 3.393 (21) x 2 | S11-S21 | 3.360 (19) x 2 |  |
| S12-S22 | 3.322 (19) x 2 | S12-S22 | 3.322 (19) x 2 | S14-S24 | 3.330 (19) x 2 | S14-S24 | 3.330 (19) x 2 |  |
| S11-S22 | 3.559 (18) x 2 | S12-S23 | 3.522 (21) x 2 | S13-S24 | 3.467 (19) x 2 | S14-S21 | 3.420 (20) x 2 |  |
| S12-S21 | 3.248 (20) x 2 | S13-S22 | 3.405 (20) x 2 | S14-S23 | 3.324 (21) x 2 | S11-S24 | 3.284 (17) x 2 |  |
| S21-S22 | 3.402 (19) x 2 | S23-S22 | 3.382 (20) x 2 | S23-S24 | 3.350 (19) x 2 | S21-S24 | 3.390 (19) x 2 |  |
| average | 3.3618 |  | 3.3854 |  | 3.3635 |  | 3.3489 |  |
| s | 0.1076 |  | 0.0783 |  | 0.0605 |  | 0.0571 |  |

| 5 K | | | | | | | | |
|---|---|---|---|---|---|---|---|---|
| V1-V2 | 2.894 (13) | | | | | | | |
| V2-V3 | 2.958 (12) | | | | | | | |
| V3-V4 | 2.750 (13) | | | | | | | |
| V4-V1 | 2.798 (14) | | | | | | | |
| average | 2.8500 | | | | | | | |
| s | 0.0935 | | | | | | | |
| | vanadium 1 | | vanadium 2 | | vanadium 3 | | vanadium 4 | |
| V1-S12 | 2.683 (20) | V2-S13 | 2.712 (20) | V3-S14 | 2.578 (17) | V4-S11 | 2.608 (17) |
| V1-S11 | 2.213 (16) | V2-S12 | 2.190 (19) | V3-S13 | 2.182 (19) | V4-S14 | 2.148 (18) |
| V1-S22 | 2.393 (15) x 2 | V2-S23 | 2.319 (14) x 2 | V3-S24 | 2.300 (13) x 2 | V4-S21 | 2.364 (16) x 2 |
| V1-S21 | 2.405 (16) x 2 | V2-S22 | 2.412 (16) x 2 | V3-S23 | 2.427 (14) x 2 | V4-S24 | 2.426 (14) x 2 |
| average | 2.4233 | | 2.4083 | | 2.3717 | | 2.3868 | |
| s | 0.1941 | | 0.2221 | | 0.1699 | | 0.1896 | |
| S21-S21 | 3.286 (23) | S23-S23 | 3.184 (16) | S23-S23 | 3.184 (16) | S21-S21 | 3.286 (23) |
| S22-S22 | 3.327 (21) | S22-S22 | 3.327 (21) | S24-S24 | 3.402 (16) | S24-S24 | 3.402 (16) |
| S11-S21 | 3.404 (19) x 2 | S13-S23 | 3.322 (22) x 2 | S13-S23 | 3.322 (22) x 2 | S11-S21 | 3.404 (19) x 2 |
| S12-S22 | 3.360 (21) x 2 | S12-S22 | 3.360 (21) x 2 | S14-S24 | 3.301 (16) x 2 | S14-S24 | 3.301 (16) x 2 |
| S11-S22 | 3.589 (16) x 2 | S12-S23 | 3.488 (18) x 2 | S13-S24 | 3.440 (17) x 2 | S14-S21 | 3.456 (20) x 2 |
| S12-S21 | 3.278 (19) x 2 | S13-S22 | 3.395 (17) x 2 | S14-S23 | 3.286 (19) x 2 | S11-S24 | 3.298 (16) x 2 |
| S21-S22 | 3.400 (18) x 2 | S23-S22 | 3.378 (18) x 2 | S23-S24 | 3.340 (18) x 2 | S21-S24 | 3.394 (18) x 2 |
| average | 3.3895 | | 3.3664 | | 3.3304 | | 3.3663 | |
| s | 0.1050 | | 0.0800 | | 0.0718 | | 0.0646 | |

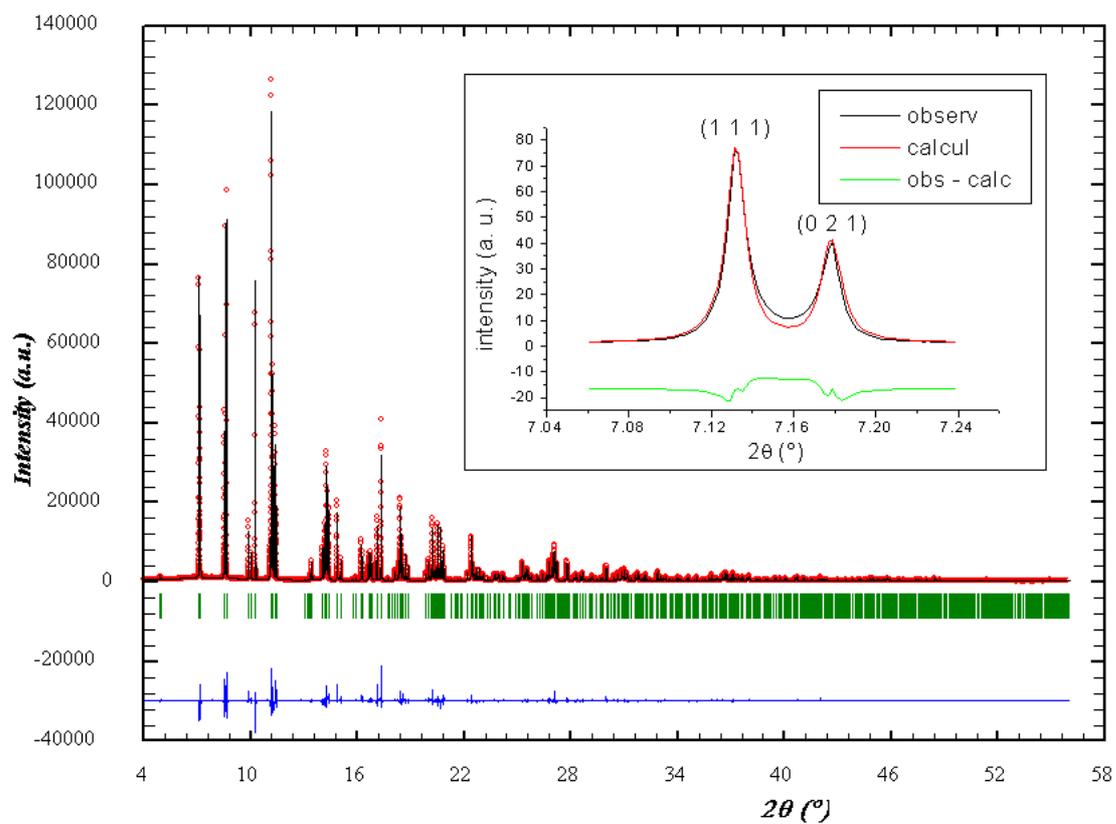

Figure 1

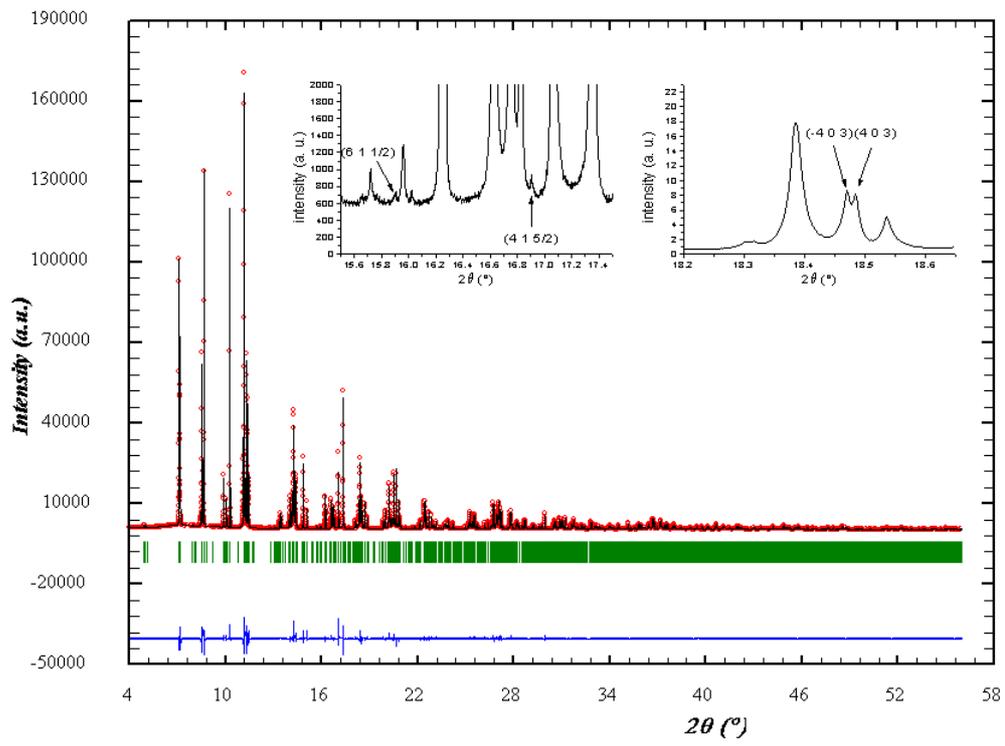

Figure 2

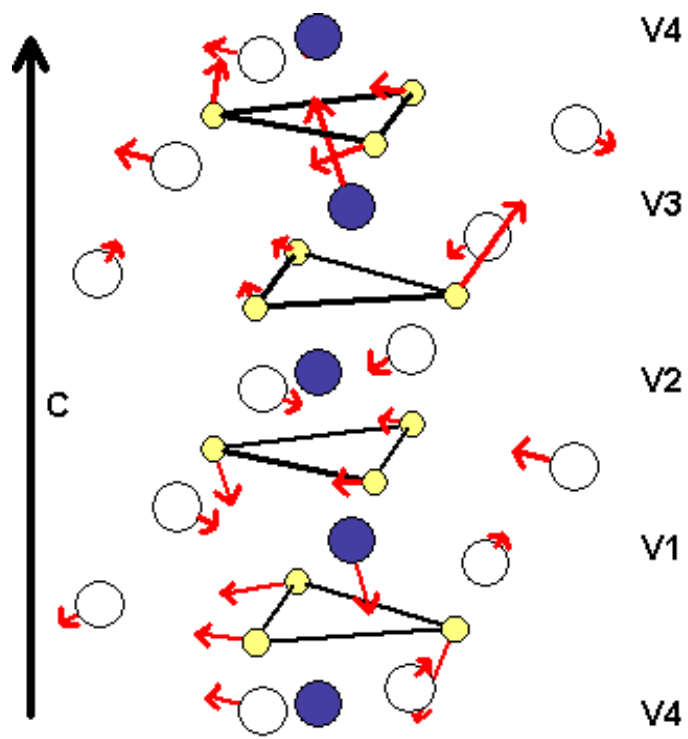

Figure 3 a

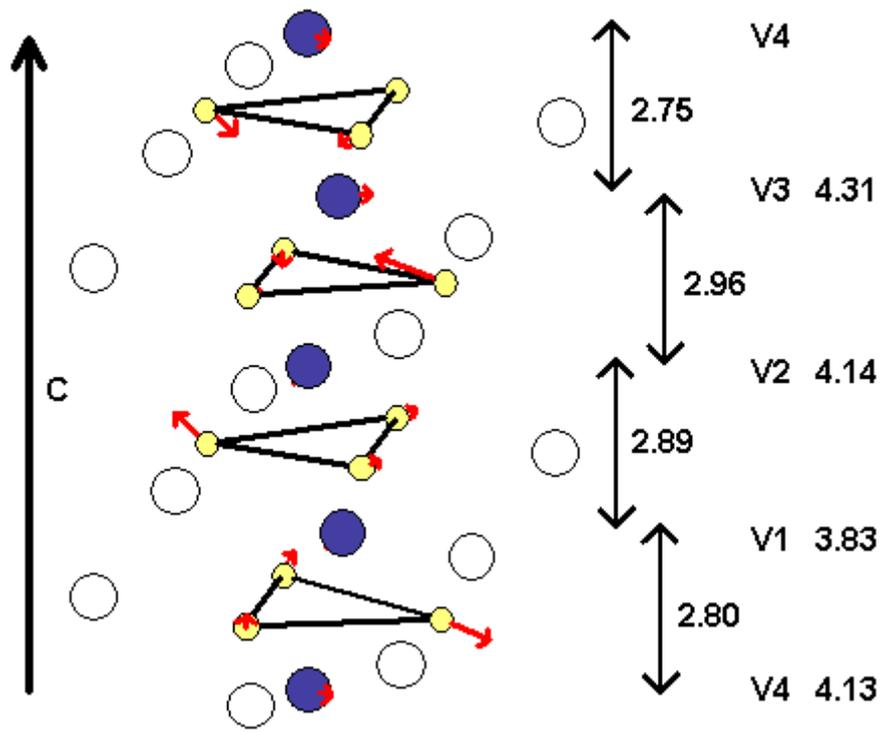

Figure 3 b